\documentclass[apj]{emulateapj-rtx4}
\usepackage{natbib}
\usepackage{graphicx}
\usepackage{threeparttable,lscape}

\shorttitle{H$_{\alpha}$ Spectral diversity of type II supernovae}
\shortauthors{Guti\'errez et al.}

\begin{document}

\title{H$_{\alpha}$ Spectral diversity of type II supernovae: correlations with photometric properties\footnote{T\lowercase{his 
paper includes data gathered with with the 6.5 m} M\lowercase{agellan} 
T\lowercase{elescopes located at} L\lowercase{as} C\lowercase{ampanas} O\lowercase{bservatory,} 
C\lowercase{hile; and the} G\lowercase{emini} O\lowercase{bservatory,} C\lowercase{erro} P\lowercase{achon,}
C\lowercase{hile} (G\lowercase{emini} P\lowercase{rogram} GS-2008B−Q−56). B\lowercase{ased on
observations collected at the} E\lowercase{uropean} O\lowercase{rganisation for} A\lowercase{stronomical}
R\lowercase{esearch in the} S\lowercase{outhern} H\lowercase{emisphere,} C\lowercase{hile} 
(ESO P\lowercase{rogrammes} 076.A-0156, 078.D-0048, 080.A-0516, \lowercase{and} 082.A-0526}}

\author{Claudia P. Guti\'errez\altaffilmark{1,2}
Joseph P. Anderson\altaffilmark{2,3},
Mario Hamuy\altaffilmark{2,1},
Santiago Gonz\'alez-Gait\'an\altaffilmark{1,2},
Gast\'on Folatelli\altaffilmark{4},
Nidia I. Morrell\altaffilmark{5},
Maximilian D. Stritzinger\altaffilmark{6},
Mark M. Phillips\altaffilmark{5},
Patrick McCarthy\altaffilmark{7},
Nicholas B. Suntzeff\altaffilmark{8},
Joanna Thomas-Osip\altaffilmark{5}}

\altaffiltext{1}{Millennium Institute of Astrophysics, Casilla 36-D, Santiago, Chile }
\altaffiltext{2}{Departamento de Astronom\'ia, Universidad de Chile, Casilla 36-D, 
Santiago, Chile}
\altaffiltext{3}{European Southern Observatory, Alonso de C\'ordova 3107, Casilla 19, Santiago, Chile}
\altaffiltext{4}{Institute for the Physics and Mathematics of the Universe (IPMU),
University of Tokyo, 5-1-5 Kashiwanoha, Kashiwa, Chiba 277-8583, Japan}
\altaffiltext{5}{Carnegie Observatories, Las Campanas Observatory, 
Casilla 601, La Serena, Chile}
\altaffiltext{6}{Department of Physics and Astronomy, 
Aarhus University, Ny
Munkegade 120, DK-8000 Aarhus C, Denmark}
\altaffiltext{7}{Observatories of the Carnegie Institution
for Science, Pasadena, CA 91101, USA}
\altaffiltext{8}{George P. and Cynthia Woods Mitchell
Institute for Fundamental Physics and Astronomy, Department of Physics and Astronomy, 
Texas A\&M University, College Station, TX 77843, USA}
\email{cgutierr@das.uchile.cl}

\begin{abstract}
We present a spectroscopic analysis of the H$_{\alpha}$ profiles of 
hydrogen-rich type II supernovae. A total of 52 type II supernovae having 
well sampled optical light curves and spectral sequences were analyzed.
Concentrating on the H$_{\alpha}$ P-Cygni profile we measure its velocity 
from the FWHM of emission and the ratio of absorption to emission ($a/e$) at
a common epoch at the start of the recombination phase,
and search for correlations between these spectral parameters and 
photometric properties of the $V$-band light curves. Testing the 
strength of various correlations we find that $a/e$ appears to be 
the dominant spectral parameter in terms of describing the diversity in our
measured supernova properties. It is found that supernovae with 
smaller $a/e$ have higher H$_{\alpha}$ velocities, more rapidly 
declining light curves from maximum, during the plateau and 
radioactive tail phase, are brighter at maximum light and have shorter
optically thick phase durations. We discuss possible explanations 
of these results in terms of physical properties of type II supernovae, 
speculating that the most likely parameters which influence the 
morphologies of H$_{\alpha}$ profiles are the mass and density profile of
the hydrogen envelope, together with additional emission components
due to circumstellar interaction.\\

\end{abstract}


\section{Introduction}

Type II Supernovae (SNe II) are produced by the final explosion of
massive ($>8$ M$_\odot$) stars. They retain a significant part of 
their hydrogen envelope at the time of the explosion, and hence their
spectra show strong Balmer lines. Studies of the variety of SNe~II
have relied on photometric analysis, cataloging this group in two 
sub-classes according to the shape of the light curve: SNe with a 
plateau (quasi-constant luminosity for a period of a few months) are
classified as SNe IIP, while SNe with steeper declining linear
light curves as SNe IIL \citep{Barbon79}. 
However, despite the role played by 
SNe~II in stellar evolution, the impact on their environments and 
their importance as standardized candles, an overall picture 
describing the physics which underpins their diversity is lacking. 
It has been suggested that SNe~IIL are produced by progenitors which
explode with smaller mass H envelopes, which then lead to SNe with
more linearly declining light curves and shorter or non-existent
`plateaus' \citep{Popov93}. Indeed, this was argued to 
be the case for the prototype SN~IIL 1979C \citep{Branch81}.
This would imply that SNe IIL progenitors suffer from a higher level of
mass-loss than their IIP counterparts. In addition, a number of 
SNe~IIL have shown evidence for circumstellar (CSM) interaction at 
late times (e.g.  SN 1986E, \citealt{Cappellaro95a};
SN~1979C, \citealt{Milisavljevic09}), which has been interpreted as 
evidence of interaction of the ejecta with the pre-supernova CSM
(see e.g. \citealt{Sahu06}, \citealt{Inserra13}). 
However, a number of authors have also claimed evidence for signs of CSM 
interaction in SNe~IIP (e.g. SN~1999em, \citealt{Pooley02};
SN~2004et, \citealt{Kotak09}; SN~2007od, 
\citealt{Inserra11}, \citealt{Andrews10};
SN~2009bw, \citealt{Inserra12b}). \\
\indent In recent years many individual studies have been published focusing
on particular properties of individual SNe, but few statistical 
studies where the spectral and photometric properties have been 
directly related are available. \citet{Patat94} found
correlations and anti-correlations between the maximum B-band magnitude
($M_{max}^B$), the color at maximum ($(B-V)_{max}$) and the ratio of
absorption to emission ($e/a$) in H$_{\alpha}$, concluding that SNe IIL 
have shallower P-Cygni profiles (larger $e/a$ values) than SNe IIP.  
\citet{Hamuy02L} analysed 17 SNe IIP and found
that SNe with brighter plateaus have higher expansion velocities.
Similar results were found by \citet{Pastorello04} with four SNe II,
who concluded that low lumino\-sity SNe have narrow spectral lines indicating 
low expansion velocities.
\citet{Hamuy03} used observations together with the analytical
models of \citet{Litvinova83, Litvinova85} to
derive physical SN~IIP properties. He 
found that more massive progenitors produce more
energetic explosions and in turn produce more nickel. These results were  
confirmed by \citet{Pastorello03} with a heterogeneous group 
of SNe~II that share a very wide range of physical properties.\\
\indent Despite the above results, it is currently unclear whether underlying spectral and 
photometric relations exist for the whole ensemble of SN~II events.
Therefore, here we attempt to remedy this situation by presenting
an initial statistical analysis of various spectroscopic and photometric properties
of a large sample of SNe~II.\\
\indent In this letter we present results showing the diversity of H$_{\alpha}$ 
P-Cygni profiles, and relations between spectral and photometric 
parameters for a sample of 52 SNe. The letter is organized as follows.
In \S\ 2 we outline our SN sample and we define the 
measurements, then in \S\ 3 we present the results. In \S\ 4 
possible physical explanations of those results are discussed, and finally 
in \S\ 5 we list our conclusions. We note that a detailed analysis
of the $V$-band light curve properties of the currently analyzed sample of SN~II 
is being presented in Anderson et al. (submitted, hereafter A14).

\section{SN II data and measurements}

The sample of SNe II employed in this study was obtained by the Carnegie
Supernova Project (CSP, \citealt{Hamuy06}) between 2004 and 
2009 plus previous campaings: the Calan/Tololo Supernova Survey (CT),
the Cerro Tololo SN program, the Supernova Optical and Infrared Survey 
(SOIRS) and the Carnegie Type II Supernova Survey (CATS). The full spectroscopic
sample will be published in an upcoming paper. Data reductions were
performed with IRAF\footnote{IRAF is distributed by the National Optical 
Astronomy Observatories (NOAO), which are operated by the Association of 
Universities for Research in Astronomy (AURA), Inc., under cooperative 
agreement with the National Science Foundation.} using the standard 
routines (bias subtracted, flat-field correction, 1-D extraction and 
wavelength correction). Detailed discussion of
spectroscopic observations and reductions for CSP was first presented in 
\citet{Hamuy06}, then outlined further in 
\citet{Folatelli13}. These are also applicable to previous data.
From this database we selected a sub-sample of events with sufficient
data to measure our spectral and photometric parameters. 
SN~IIn and SN~IIb were not analysed in this work.\\
\indent SNe II show a large diversity in their spectra. As the dominant
spectral feature is the H$_{\alpha}$ P-cygni profile,
for this initial study we concentrate on this line's  properties.
The H$_{\alpha}$ line presents a diversity that can 
be derived from the shape and strength in the emission and absorption,
and in the line width. Figure~1 shows the variety in SNe~II 
H$_{\alpha}$ P-Cygni profiles, where the SNe are ordered in terms of 
an increasing ratio of absorption to emission ($a/e$) components (as 
defined below) around a common epoch at the start in the recombination phase.
We see that the absorption is the component which 
changes most from one SN to another rather than the emission.
There are SNe with little absorption (e.g. SN~2006ai,
SN~2006Y), while there are others with boxy absorption profiles 
(e.g. SN~2003cx, SN 2007X). One can observe in Figure~1 that the first 
SNe show little absorption compared to emission. Gradually
the SNe change to show more classic P-Cygni profiles with significant
absorption components. A number of SNe show an 
extra absorption component on the blue side of H$_{\alpha}$
(e.g. SN~2003hn, SN~2007od, SN~2008aw).

\begin{figure*}
\epsscale{1}
\plotone{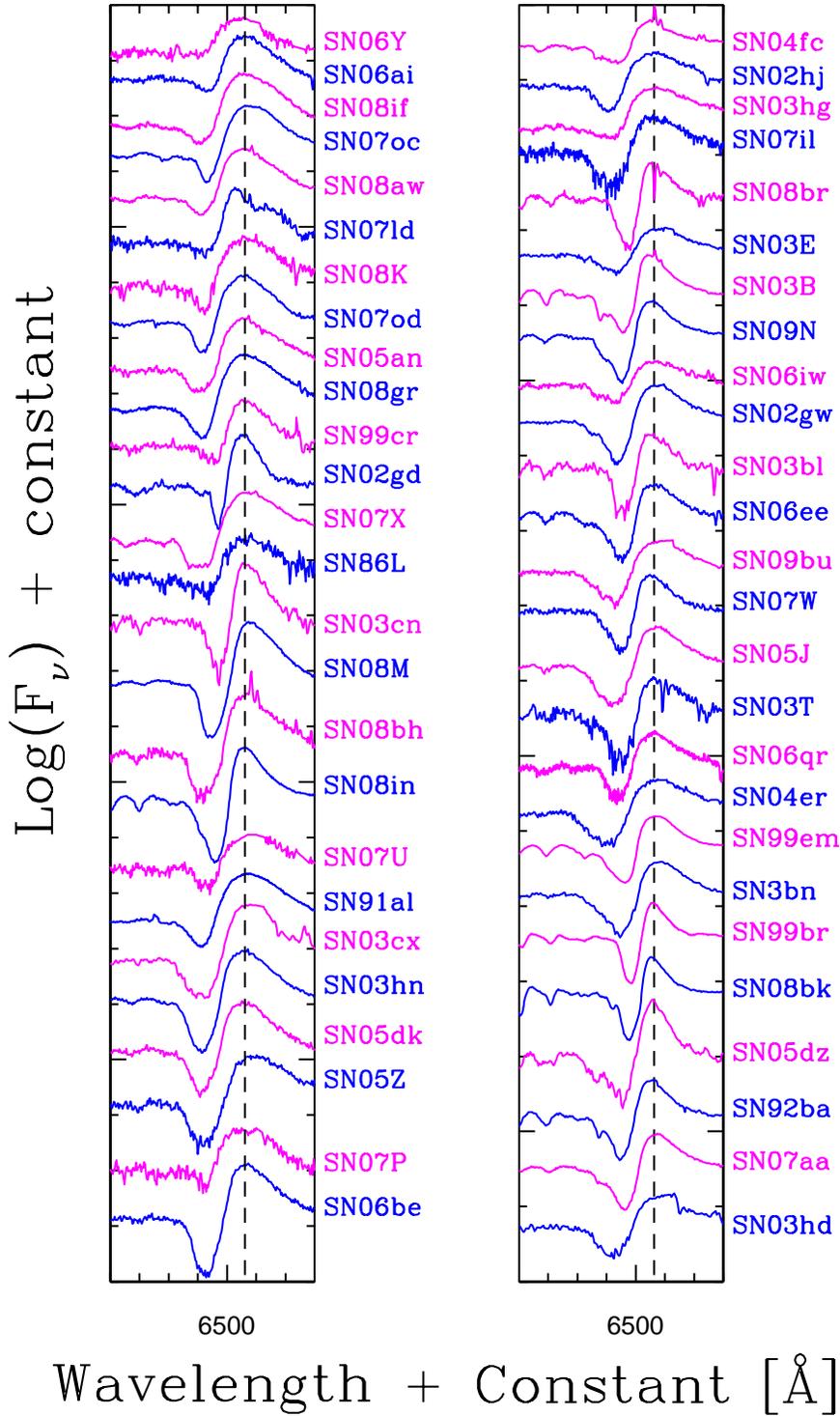}
\caption{Variety in SN II $H_{\alpha}$ P-Cygni profiles
ordered in terms of increasing $a/e$ starting top left, finishing bottom right.
The Host galaxy features were removed  and the spectra are shifted to be
centered on the peak of H$_{\alpha}$ emission.
The epoch of the spectra shown are those in closest time proximity
to $t_{tran}+10$. In general the difference between epochs of the spectra and
$t_{tran}+10$ is within $\pm10$ days. }
\end{figure*}

\indent To analyze the SNe spectra within our sample we define two measurements: 
\textbf{(1)} the expansion velocity in H$_{\alpha}$ via the FWHM of 
the emission, and \textbf{(2)} the ratio of equivalent widths of
absorption to emission $(a/e)$ components of H$_{\alpha}$. This ratio 
was initially proposed by \citet{Patat94} as the flux ratio
of the emission to absorption. However, we choose $a/e$ because in a 
few SNe H$_{\alpha}$ shows an extremely weak absorption component. In 
order to relate spectral and light curve properties we use the 
\textit{V}-band photometric properties as defined by A14:
\textit{$s_{1}$}: initial decline from the maximum (magnitudes $100d^{-1}$),
\textit{$s_{2}$}: `plateau' decline rate (magnitudes $100d^{-1}$), 
\textit{$s_{3}$}: radioactive tail decline (magnitudes $100d^{-1}$), 
\textit{M$_{max}$}: magnitude at \textit{V}-band maximum, and 
\textit{OPTd}: optically thick phase duration (days): time from the explosion epoch
through to the end of the plateau phase.
We define a common epoch in order to measure spectral properties,
which we identified in the light curves: the \textit{B}-band transition time
plus 10 days ($t_{tran}+10$, at the start of the recombination phase).
The transition time is defined as the transition between 
\textit{$s_{1}$} and \textit{$s_{2}$} determined by chi-square minimization.
It is measured in the \textit{B}-band because the transition is more 
evident than in \textit{V}-band, and therefore we can include more SNe in our analysis.
We interpolate all the spectral measurements to this epoch.
These parameters are all labeled in the light-curve parameter
schematic presented in Figure~2 (left).

\begin{figure*}
\includegraphics[width=9.2cm]{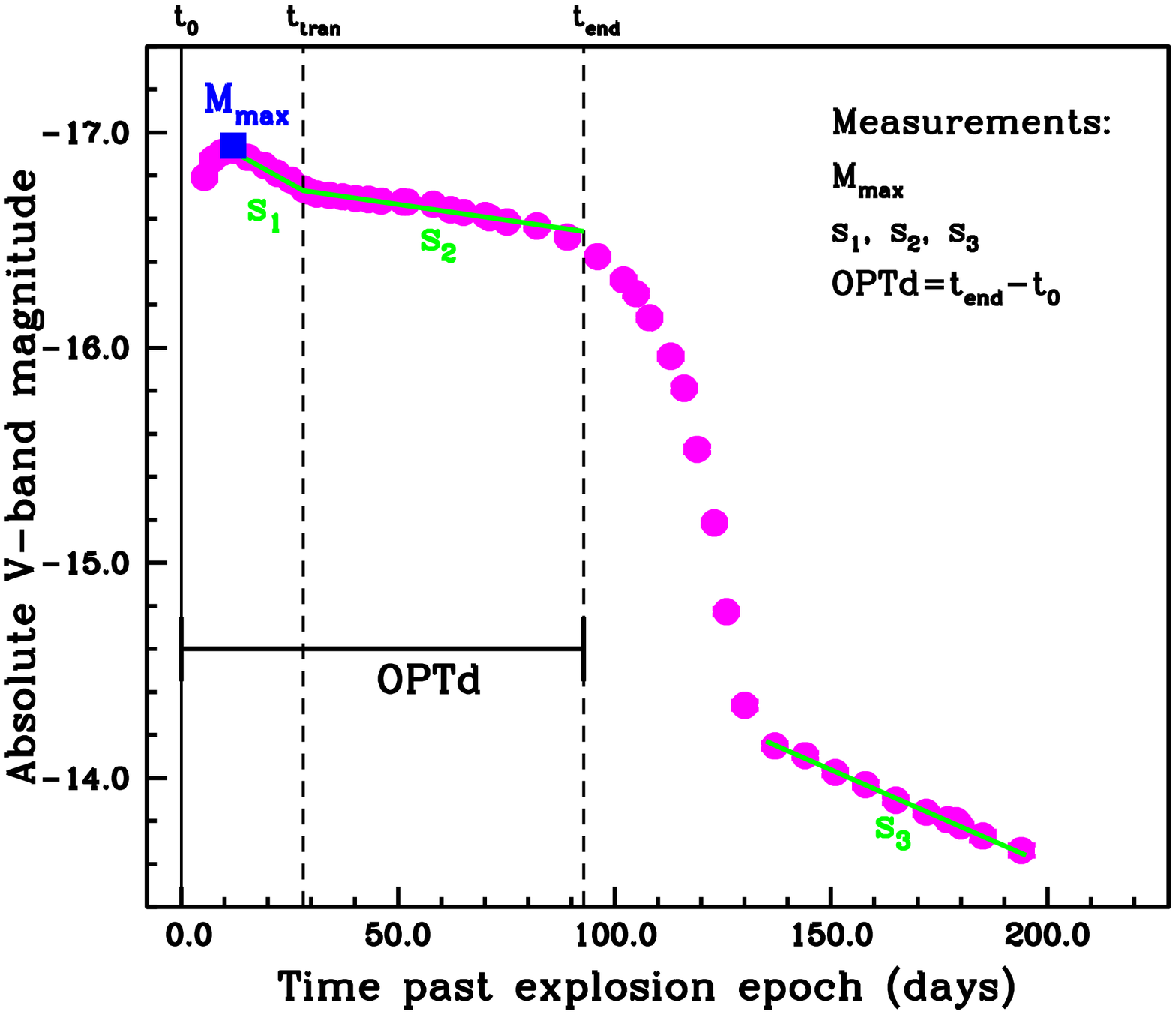}
\includegraphics[width=9.2cm]{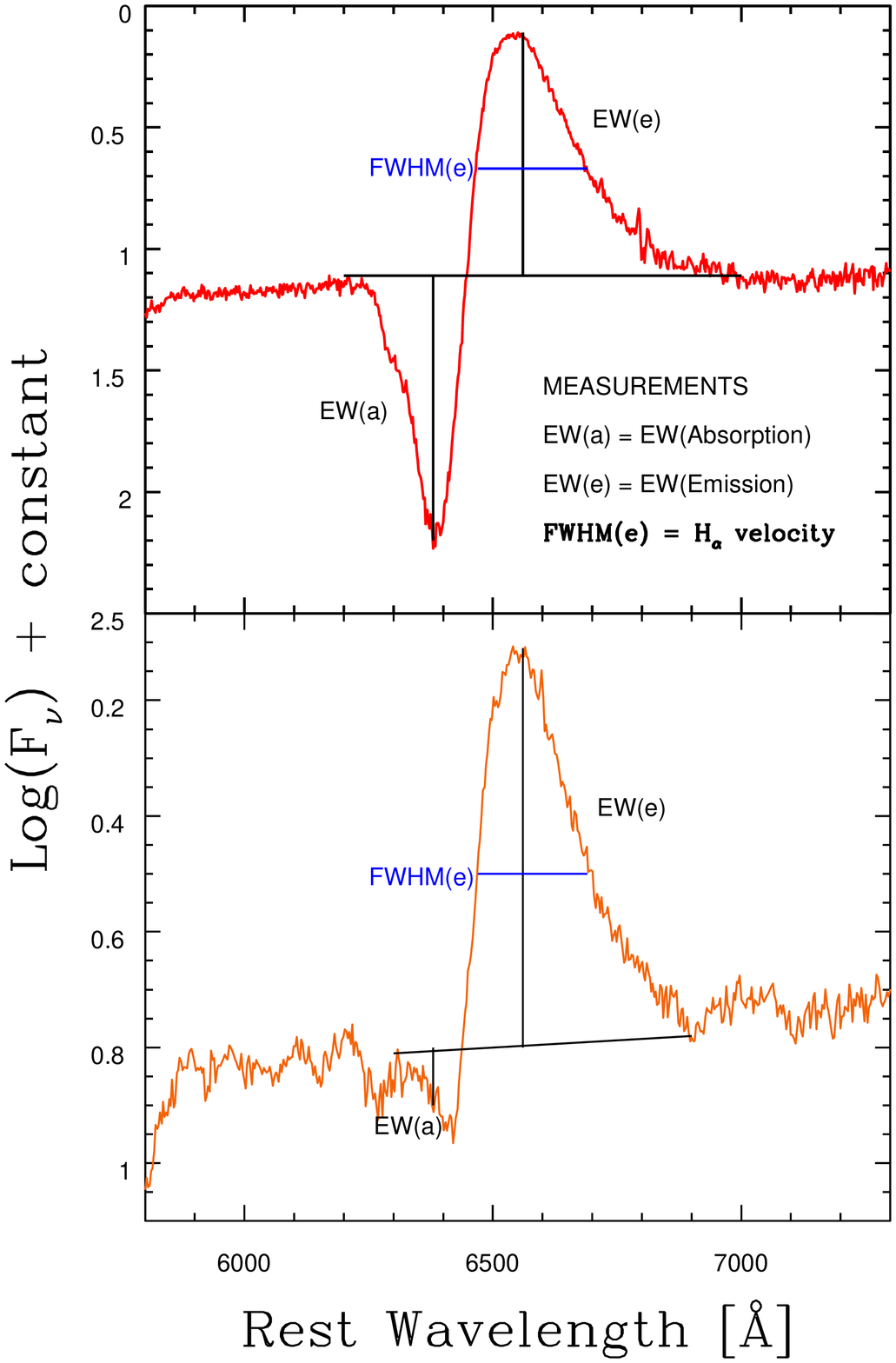}
\caption{\textit{Left:} An example of the light-curve parameters measured for each SN
within the sample in the \textit{V}-band (note that while on the figure $t_{tran}$ is shown 
with respect \textit{V}-band data, our analysis uses this epoch as defined in the \textit{B}-band
as this allows to include more SNe). Observed absolute magnitude at peak, \textit{M$_{max}$} is shown in blue, 
as applied to the dummy data points (magenta) of a SN IIP. The positions of the three 
measured slopes; \textbf{$s_1$}, \textbf{$s_2$}, and \textbf{$s_3$} are shown in
green. The optically thick phase duration, \textit{OPTd} is indicated in black.
Three time epochs are labeled: $t_0$, the explosion epoch; $t_{tran}$, the transition 
from \textbf{$s_1$} to \textbf{$s_2$}; and $t_{end}$, the end of the optically thick phase.
\textit{Right:} Examples of spectral measurements. Top: SN~2009bz shows a normal
H$_{\alpha}$ P-Cygni profile with absorption and emission components 
well defined. Bottom: SN~2008aw shows a peculiar profile with a extra
component on the blue side. In both plots, the blue line represents the
FWHM of emission, which is used for the velocities values, and black 
horizontal line defines the continuum level used to measure the EW of 
emission and absorption.}
\end{figure*}

\indent To estimate SN ejecta expansion velocities through H$_{\alpha}$ the 
minimum flux of the absorption component of P-Cygni line profile
is commonly used. However, a few SNe in this sample present an extremely
weak absorption component, complicating this method. Therefore,
we employ the FWHM of emission line for velocity estimations. To 
verify the concordance between these methods we measure velocities
from the minima of absorption and the FWHM of the emission in
H$_{\alpha}$ in SNe with a well defined absorption, finding consistent
results. The ratio of absorption to emission ($a/e$) in H$_{\alpha}$
was obtained by measuring the equivalent widths (EW) of each component.
Examples of these measurements are shown in Figure~2 (right). The top panel 
shows a normal H$_{\alpha}$ P-Cygni profile, i.e., a profile with 
well defined absorption and emission components, while the bottom panel 
shows a peculiar profile with an extra absorption component on the blue
side. Similar features were identified by \citet{Leonard01},
\citet{Leonard02a} and \citet{Leonard02b} in SN~1999em
as high velocity (HV) features, while in the case of SN~2005cs the line 
was identified as Si II $\lambda6355$ absorption \citep{Pastorello06}. 
This peculiar structure complicates
measurements of the EW of absorption, because it is hard to objectively 
define the continuum. Therefore, we simply trace a straight line along the 
absorption feature to mimic the continuum flux, which can be seen in Figure~2 (right).
All spectral measurements were performed with IRAF using the \textit{splot} package.
The errors for the H$_{\alpha}$ velocity and $a/e$ are mainly dominated
by how the continuum is defined. Errors were obtained by measuring many
times the FWHM and the EW, respectively, changing the trace of the 
continuum. Using these multiple measurements we calculate a mean and 
take the standard deviation to be the error on that measurement.

\section{Results}

In Table~1 we list the measured spectral and photometric parameters:
H$_{\alpha}$ velocity, $a/e$, \textit{$s_{1}$}, \textit{$s_{2}$}, 
\textit{$s_{3}$}, M$_{max}$ and \textit{OPTd} for each SN, together with the host
galaxy, the heliocentric radial velocity and $t_{tran}$. We searched for correlations 
between all seven of our defined parameters against each other at different 
epochs: $t_{tran}$, $t_{tran}-10$, $t_{tran}+10$, $t_{tran}+20$, $t_{tran}+30$ (measured
in \textit{B}-band), and 30 and 50d since explosion. Using the Pearson 
correlation test, $a/e$ was observed to be the dominant measured spectral parameter as 
it has the highest correlation with all other parameters at all epochs. However, at
$t_{tran}+10$ the correlations are strongest, hence this time was chosen as the common epoch.
This is justified from a physical point of view because at this epoch all SNe are entering to a 
similar phase in their evolution, i.e. the recombination phase.
The photometric parameter \textit{$s_{1}$} has the highest mean correlation, 
however it shows no correlation with the H$_{\alpha}$ velocity,  
while \textit{$s_{2}$} shows high correlation with all parameters.
In A14 the photometric correlations are presented. Table~2 shows 
the strength of the correlations between all our parameters, plus the number of events
(within each correlation), and the probability of finding such a correction by chance.
The light curve parameters plus the H$_{\alpha}$ velocity are plotted
versus $a/e$ in Figure~3. The plot shows that SNe with smaller $a/e$ 
have higher H$_{\alpha}$ velocities,  more rapidly declining light curves 
after maximum, both in the `plateau' and radioactive tail phases, are 
brighter and have shorter \textit{OPTd} values. SNe with higher $a/e$ show opposite
behavior. Given that \textit{OPTd} and \textit{$s_{3}$} are most likely related to
the envelope/ejecta mass (see A14 for detailed discussion), 
this would appear to imply that $a/e$ is also related to the mass retained by the
progenitor before explosion. Indeed, this further points to SNe historically 
classified as IIL (high \textit{$s_{2}$}) having smaller mass envelopes at the 
epoch of explosion than their IIP (low \textit{$s_{2}$}) counterparts.
Moreover, we see a continuum of events in terms of spectral
diversity, thus suggesting a possible continuum in pre-SN envelope masses.\\
\indent In Figure~3 one can see an extreme object specifically in panels A, B and E,
SN~2006Y. To test whether this object drives our correlations we re-do the Pearson test for
all correlations removing this object. While of the strength of the correlations decrease
very slightly all correlations hold, and therefore this object is not driving our results.\\
\indent We also searched for correlations between $a/e$ and the photospheric velocity,
derived from the minimum of absorption of the Fe II $\lambda$5169 $\AA$ line at
the epoch defined above. However, perhaps surprisingly, no evidence for
correlation was found.

\begin{figure*}
\epsscale{1.1}
\plotone{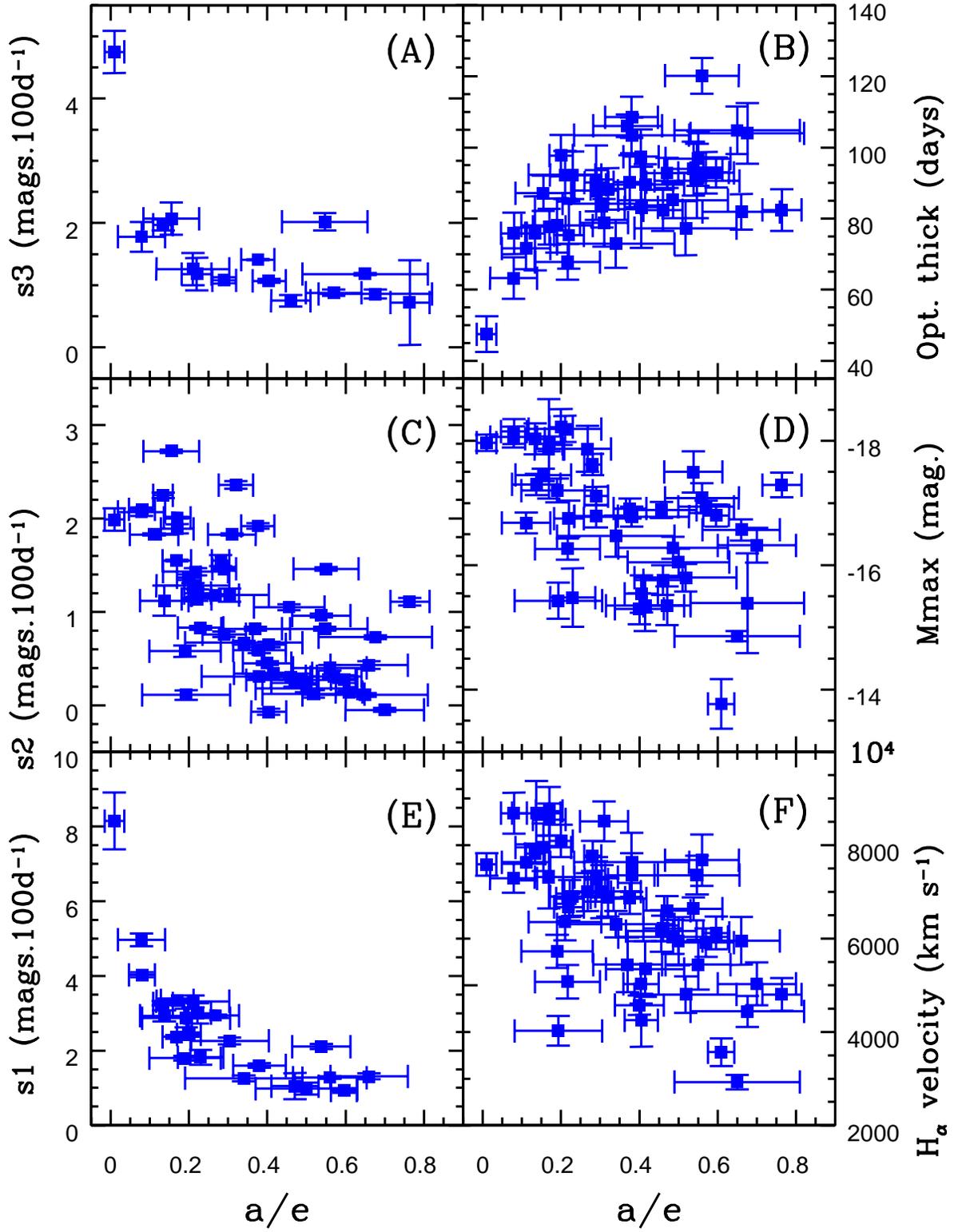}
\caption{Relations between $a/e$, H$_{\alpha}$ velocity, \textit{$s_{1}$}, 
\textit{$s_{2}$}, \textit{$s_{3}$} and M$_{max}$ at $t_{tran}+10$. 
{\it Panel A:} $a/e$ vs. \textit{$s_{3}$}. 
{\it Panel B:} $a/e$ vs. Optically thick phase. 
{\it Panel C:} $a/e$ vs. \textit{$s_{2}$}.
{\it Panel D:} $a/e$ vs. M$_{max}$.
{\it Panel E:} $a/e$ vs. \textit{$s_{1}$}.
{\it Panel F:} $a/e$ vs. H$_{\alpha}$ velocity.}
\end{figure*}

\section{Discussion}

We have presented and analysed the H$_{\alpha}$
spectral diversity in 52 SNe II and their correlations with 
photometric parameters. Analyzing the sample we see a variety in
the H$_{\alpha}$ P-Cygni profiles which can be derived from the shape 
and strength in the emission and absorption, and in the line width.
\citet{Patat94} found that M$_{max}^B$ and $(B-V)_{max}$ 
correlate with $e/a$ in H$_{\alpha}$, concluding that SNe~IIL have 
larger $e/a$ values (i.e. small $a/e$ values). While in our sample we 
have not made distinctive IIP-IIL classification, our results are 
consistent with these of \citet{Patat94}, which show that SNe with 
high \textit{$s_{2}$} values (faster declining light
curves) have small $a/e$ values and are more luminous.\\
\indent \citet{Arcavi12} identified a subdivision of SNe~II 
(based on 21 events in the \textit{R}-band) suggesting that
SNe~IIL and SNe~IIP are not members of one continuous class and
may result from different physical progenitor systems.
However, A14 with a bigger sample (116 events in the \textit{V}-band) 
suggest an observational continuum of events which may be driven by differences
of envelope mass at the epoch of explosion, 
a parameter which is most directly constrained in A14 through observations
of the optically thick phase duration (\textit{OPTd}) and the decline rate
during the radioactive tail (\textit{$s_{3}$}). 
This conclusion of an observational continuum is also
supported by the spectral analysis presented in this paper (see Figure~3), where
differences in the spectral parameters (especially $a/e$) may also be explained 
by changes in the hydrogen envelope mass retained.\\
\indent \citet{Schlegel96} discussed possible explanations for 
the behavior of the H$_{\alpha}$ P-Cygni profile with the most likely
being: \textbf{(1)} extra emission fills in the absorption component 
(as can be seen in SN~2008aw, Figure~2, bottom); \textbf{(2)} the 
envelope mass is low; and \textbf{(3)} a steep density gradient
in the hydrogen envelope. The first explanation invokes scattering 
of emission off either CSM or the outer envelope (in the case of very
extended envelope). The second 
explanation is described in a low-mass envelope, where there is less 
absorbing material, so little P-Cygni absorption component will be formed.
The third explanation argues that a very steep density gradient 
implies less absorbing material at high velocities, and so does not 
produce a well defined P-Cygni profile. Although these considerations
could explain the diversity found in our sample, numerous studies 
discuss other explanations based on the complex P-Cygni
line profiles. \citet{Baron00} granted the term 
`complicated P-Cygni profile' to explain the double P-Cygni absorption
found in Balmer Series and He I $\lambda5876$ in SN~1999em,
concluding that these absorption features arise in two velocity 
regions in the expanding ejecta of the SN at different velocities.
\citet{Pooley02} argue that this extra component 
might be the signature of weak interaction with a low density CSM,
while \citet{Chugai07} attributes these features to ejecta
wind interactions.
In conclusion, the change in H$_{\alpha}$ P-Cygni profile ($a/e$ and FWHM of emission) 
is most likely related to two parameters: changes in the envelope properties (i.e. its 
mass and density profile) and the degree of CSM interaction.
Although the possible explanations for the behavior of the
H$_{\alpha}$ P-Cygni profile have been exposed, these 
extra components could be attributed to HV H I features or, absorption
lines of other ions (Si II). This issue will be further explored after 
a full spectral analysis. This will determine if similar features are also present in 
the blue side of He I $\lambda5876$ and H$_{\beta}$.

\section{Conclusions}

We have presented an initial analysis of the spectral diversity H$_{\alpha}$
of SNe~II and how this relates to light curve properties. It has been found 
that while much diversity and peculiarities exist, spectral and photometric 
properties do appear to be correlated which can be linked to pre-SN properties. 
We finally list our main conclusions:
\begin{itemize}

\item $a/e$ is an important parameter describing the spectral diversity of 
SNe~II.

\item SNe with low $a/e$ values appear to have high $H_{\alpha}$ velocities
and decline rates, are brighter and have a smaller \textit{OPTd} values.

\item While any definitive spectral distinction between IIP and IIL is not clear,
SNe with higher \textit{$s_{2}$} values (i.e. more `linear' SNe)
have smaller $a/e$ values, have higher H$_{\alpha}$ velocities, and are more luminous.

\item We speculate that the envelope mass retained before explosion 
and the density gradient play a very important role to determine the 
differences of H$_{\alpha}$ P-Cygni 
profile.

\item CSM interaction could also be a cause of the change in the 
P-Cygni profiles, suggesting that faster declining SNe have more intense
interactions.

\end{itemize}

This paper presented a first analysis of SN~II spectral from CSP. The full 
analysis of that sample (optical and near IR photometry and spectroscopy) 
promises to significantly further our knowledge of the SN~II phenomenon.

\acknowledgments
We thank the annonymous referee for their useful suggestions.
C.P.G. acknowledges support from CONICYT-AGCI PhD studentship.
C.P.G., J.P.A., M.H., S.G., acknowledge support by projects IC120009 ``Millennium
Institute of Astrophysics (MAS)" and P10-064-F ``Millennium Center for Supernova
Science" of the Iniciativa Cient\'ifica Milenio del Ministerio Econom\'ia,
Fomento y Turismo de Chile.
J.P.A. acknowledges support by CONICYT through
FONDECYT grant 3110142. S.G. aknowledgeds support by CONICYT through FONDECYT 
grant 3130680. M.S.  gratefully  acknowledges the generous support provided by 
the Danish Agency for Science and Technology and Innovation  realized through a 
Sapere Aude Level 2 grant.
This research has made use of the NASA/IPAC
Extragalactic Database (NED) which is operated by the Jet Propulsion 
Laboratory, California Institute of Technology, under contract with the
National Aeronautics.

\newpage 
\clearpage
\begin{landscape}
\begin{deluxetable}{ccccccccccccccccc}
\tiny
\centering
\setlength{\tabcolsep}{1pt} 
\tablecolumns{11} 
\tablewidth{0pc} 
\tablecaption{SNe~II spectral and photometric parameters}
\tablehead{ \colhead{SN} & \colhead{Host}  & \colhead{Recession velocity} & \colhead{$t_{tran}$} & \colhead{M$_{max}$} & \colhead{\textit{$s_{1}$}}
& \colhead{\textit{$s_{2}$}} & \colhead{\textit{$s_{3}$}} & \colhead{$Optd$} & \colhead{$a/e$} & \colhead{$H_{\alpha}$ velocity}\\
 & \colhead{galaxy} &  \colhead{(km s$^{-1}$)} & \colhead{(MJD)} & \colhead{(mag)} & \colhead{(mag $100d^{-1}$)} & \colhead{(mag $100d^{-1}$)}
 & \colhead{(mag $100d^{-1}$)} & \colhead{(days)} &  &  \colhead{(km s$^{-1}$ )}}
\startdata
1986L	& NGC 1559	&	1305	& $46747.4\pm0.5$	& $-18.19\pm0.2$ & $3.32\pm0.16$ & $1.28\pm0.03$ & \nodata        & $92.24\pm6.71$  & $0.21\pm0.09$   & $6354\pm392$    \\
1991al	& LEDA 140858	&	4575	& $48478.1\pm1.3$	& $-17.62\pm0.2$ & \nodata	 & $1.55\pm0.06$ & $1.26\pm0.26$  & \nodata	    & $0.28\pm0.02$   & $7771\pm320$    \\
1992ba	& NGC 2082	&	1185	& $48926.3\pm1.0$	& $-15.39\pm0.8$ & \nodata	 & $0.73\pm0.02$ & $0.86\pm0.07$  & $103.97\pm8.54$ & $0.68\pm0.14$   & $4439\pm334$    \\
1999br	& NGC 4900	&	960	& $51308.4\pm0.5$	& $-13.77\pm0.4$ & \nodata	 & $0.14\pm0.02$ & \nodata	  & \nodata	    & $0.61\pm0.03$   & $3566\pm297$    \\
1999cr	& ESO 576-G034	&	6069	& $51273.6\pm0.7$	& $-17.20\pm0.2$ & $1.80\pm0.06$ & $0.58\pm0.06$ & \nodata	  & $78.06\pm7.62$  & $0.19\pm0.09$   & $5728\pm357$    \\
1999em	& NGC 1637	&	717	& $51509.7\pm0.7$	& $-16.94\pm0.1$ & \nodata       & $0.31\pm0.02$ & $0.88\pm0.05$  & $92.86\pm5.83$  & $0.57\pm0.07$   & $5915\pm306$    \\
2002gd	& NGC 7537	&	2676	& $52581.6\pm0.8$	& $-15.43\pm0.3$ & $2.87\pm0.25$ & $0.11\pm0.05$ & \nodata	  & \nodata	    & $0.19\pm0.11$   & $4023\pm320$    \\
2002gw	& NGC 922	&	3084	& $52583.1\pm1.0$	& $-15.76\pm0.2$ & \nodata	 & $0.30\pm0.03$ & $0.75\pm0.09$  & $82.33\pm5.83$  & $0.46\pm0.05$   & $6217\pm274$    \\
2002hj	& NPM1G +04.0097&	7080	& $52595.8\pm3.1$	& $-16.91\pm0.2$ & \nodata	 & $1.92\pm0.03$ & $1.41\pm0.01$  & $90.24\pm7.62$  & $0.38\pm0.04$   & $6857\pm334$    \\
2003B	& NGC 1097	&	1272	& $52666.2\pm1.9$	& $-15.54\pm0.3$ & \nodata	 & $0.65\pm0.03$ & $1.07\pm0.03$  & $83.19\pm11.4$  & $0.40\pm0.04$   & $4251\pm658$    \\
2003E	& MCG -4-12-004	&	4470	& $52661.5\pm3.5$	& \nodata	 & \nodata	 & $-0.07\pm0.03$& \nodata	  & $97.42\pm7.62$  & $0.40\pm0.04$   & $5028\pm424$    \\
2003T	& UGC 4864	&	8373	& $52686.8\pm1.5$	& \nodata	 & \nodata	 & $0.82\pm0.02$ & $2.02\pm0.14$  & $90.59\pm10.44$ & $0.55\pm0.11$   & $7360\pm411$	\\
2003bl	& NGC 5374	&	4377	& $52736.7\pm1.2$	& $-15.35\pm0.3$ & $1.05\pm0.35$ & $0.24\pm0.04$ & \nodata	  & $92.81\pm4.24$  & $0.47\pm0.06$   & $6596\pm311$    \\
2003bn	& 2MASX J10023529& 	3831	& $52729.7\pm8.6$	& $-16.80\pm0.2$ & $0.93\pm0.06$ & $0.28\pm0.04$ & \nodata	  & $92.97\pm4.24$  & $0.60\pm0.03$   & $6121\pm352$    \\
2003cn	& IC 849	&	5433	& $52743.6\pm1.7$	& $-16.26\pm0.2$ & \nodata	 & $1.43\pm0.04$ & \nodata	  & $67.80\pm5.00$  & $0.22\pm0.08$   & $5074\pm361$    \\
2003cx	& NEAT J135706.53&	11100	& $52754.7\pm4.2$	& $-16.79\pm0.2$ & \nodata	 & $0.76\pm0.03$ & \nodata	  & $87.82\pm5.83$  & $0.29\pm0.05$   & $7314\pm343$    \\
2003hd	& MCG -04-05-010&	11850	& $52886.9\pm1.6$	& $-17.29\pm0.2$ & \nodata	 & $1.11\pm0.04$ & $0.72\pm0.68$  & $82.39\pm5.83$  & $0.76\pm0.05$   & $4800\pm350$	\\
2003hg	& NGC 7771	&	4281	& $52898.3\pm3.3$	& \nodata 	 & $1.60\pm0.06$ & $0.59\pm0.03$ & \nodata	  & $108.50\pm5.83$ & $0.38\pm0.07$   & $7360\pm466$	\\
2003hn	& NGC 1448	&	1170	& $52900.9\pm3.9$	& $-17.11\pm0.1$ & \nodata	 & $1.46\pm0.02$ & $1.08\pm0.05$  & $90.10\pm10.44$ & $0.29\pm0.03$   & $7268\pm375$	\\
2004er	& MCG -01-7-24	&	4411	& $53319.3\pm0.4$	& $-17.08\pm0.2$ & $1.28\pm0.03$ & $0.40\pm0.03$ & \nodata	  & $120.15\pm5.00$ & $0.56\pm0.09$   & $7680\pm553$	\\
2004fc	& NGC 701	&	1831	& $53335.7\pm0.4$	& \nodata	 & \nodata	 & $0.82\pm0.02$ & \nodata	  & $106.06\pm3.16$ & $0.37\pm0.09$   & $5440\pm585$	\\
2005an	& SO 506-G11	&	3206	& $53466.1\pm0.3$	& \nodata	 & $3.34\pm0.06$ & $1.89\pm0.05$ & \nodata	  & $77.71\pm5.00$  & $0.17\pm0.04$   & $8548\pm343$	\\
2005dk	& C 4882	&	4708	& $53638.4\pm0.9$	& \nodata	 & $2.26\pm0.09$ & $1.18\pm0.07$ & \nodata	  & $84.22\pm6.71$  & $0.30\pm0.10$   & $7008\pm567$	\\
2005dz	& GC 12717	&	5696	& $53666.7\pm0.8$	& $-16.57\pm0.2$ & $1.31\pm0.08$ & $0.43\pm0.04$ & \nodata	  & $81.86\pm5.00$  & $0.66\pm0.10$   & $5952\pm512$	\\
2005J	& NGC 4012	&	4183	& $53421.1\pm0.4$	& $-17.50\pm0.3$ & $2.11\pm0.07$ & $0.96\pm0.02$ & \nodata	  & $94.03\pm7.62$  & $0.54\pm0.07$   & $6637\pm245$	\\
2005Z	& NGC 3363	&	5766	& $53432.9\pm1.0$	& \nodata	 & \nodata	 & $1.83\pm0.01$ & \nodata	  & $78.84\pm6.71$  & $0.31\pm0.06$   & $8512\pm430$	\\
2006Y	& anon		&	10074	& $53794.7\pm0.6$	& $-17.97\pm0.13$& $8.15\pm0.76$ & $1.99\pm0.12$ & $4.75\pm0.34$  & $47.49\pm5.00$  & $0.01\pm0.02$   & $7588\pm244$	\\
2006ai	& ESO 005- G 009&	4571	& $53813.7\pm1.0$	& $-18.06\pm0.2$ & $4.97\pm0.17$ & $2.07\pm0.04$ & $1.78\pm0.24$  & $63.26\pm5.83$  & $0.08\pm0.06$   & $7291\pm307$	\\
2006be	& IC 4582	&	2145	& $53835.4\pm0.5$	& $-16.47\pm0.3$ & $1.26\pm0.08$ & $0.67\pm0.02$ & \nodata	  & $72.89\pm6.71$  & $0.34\pm0.15$   & $6308\pm283$	\\
2006ee  & NGC 774	&       4620    & $53997.2\pm1.3$	& $-16.28\pm0.2$ & \nodata	 & $0.27\pm0.02$ & \nodata  	  & $85.17\pm5.00$  & $0.49\pm0.14$   & $6034\pm366$	\\
2006iw	& 2MASX J23211915&	9226	& $54049.40\pm1.9$	& $-16.89\pm0.1$ & \nodata	 & $1.05\pm0.03$ & \nodata 	  & \nodata	    & $0.46\pm0.09$   & $6162\pm448$	\\
2006qr	& MCG -02-22-023&	4350	& $54098.2\pm1.2$	& \nodata	 & \nodata	 & $1.46\pm0.02$ & \nodata	  & $96.85\pm7.62$  & $0.55\pm0.08$   & $5440\pm535$	\\
2007aa	& NGC 4030	&	1465	& $54162.7\pm0.8$	& $-16.32\pm0.3$ & \nodata	 & $-0.05\pm0.02$& \nodata	  & \nodata	    & $0.70\pm0.10$   & $5028\pm462$	\\
2007il	& IC 1704	&	6454	& $54393.4\pm2.0$	& $-16.78\pm0.2$ & \nodata	 & $0.31\pm0.02$ & \nodata	  & $103.43\pm5$    & $0.38\pm0.15$   & $7634\pm636$	\\
2007ld	& SDSS J204929.40& 	8994	& $54402\pm1.9$ 	& $-17.30\pm0.2$ & $2.93\pm0.15$ & $1.12\pm0.16$ & \nodata	  & \nodata	    & $0.14\pm0.06$   & $8685\pm690$	\\
2007oc	& NGC 7418	&	1450	& $54419.8\pm0.4$	& $-16.68\pm0.2$ & \nodata	 & $1.83\pm0.01$ & \nodata	  & $71.62\pm5.83$  & $0.11\pm0.06$   & $7634\pm386$	\\
2007od	& UGC 12846	&	1734	& $54427.7\pm0.3$	& $-17.87\pm0.8$ & $2.37\pm0.05$ & $1.55\pm0.01$ & \nodata	  & \nodata	    & $0.17\pm0.04$   & $7314\pm672$	\\
2007P	& ESO 566-G36 	&	12224	& $54154.9\pm3.2$	& \nodata	 & \nodata	 & $2.36\pm0.04$ & \nodata	  & $88.33\pm5.83$  & $0.32\pm0.04$   & $6880\pm410$	\\
2007U	& ESO 552-65	&	7791	& $54168.8\pm1.7$	& $-17.87\pm0.4$ & $2.94\pm0.02$ & $1.18\pm0.01$ & \nodata	  & \nodata	    & $0.27\pm0.06$   & $6994\pm407$	\\
2007W	& NGC 5105	&	2902	& $54164.3\pm1.0$	& $-15.80\pm0.2$ & \nodata	 & $0.12\pm0.04$ & \nodata	  & $77.29\pm7.62$  & $0.52\pm0.13$   & $4800\pm392$	\\
2007X	& ESO 385-G32	&	2837	& $54180.8\pm0.3$	& $-18.22\pm0.3$ & $2.43\pm0.06$ & $1.37\pm0.03$ & \nodata	  & $97.71\pm5.83$  & $0.20\pm0.03$   & $8091\pm346$	\\
2008aw	& NGC 4939	&	3110	& $54563.8\pm0.6$	& $-18.03\pm0.2$ & $3.27\pm0.06$ & $2.25\pm0.03$ & $1.97\pm0.09$  & $75.83\pm10.44$ & $0.13\pm0.02$   & $7817\pm398$	\\
2008bh	& NGC 2642	&	4345	& $54593.0\pm2.6$	& \nodata	 & $3.00\pm0.27$ & $1.20\pm0.04$ & \nodata	  & \nodata	    & $0.22\pm0.03$   & $6857\pm397$	\\
2008bk	& NGC 7793	&	227.	& $54602.3\pm1.4$	& $-14.86\pm0.1$ & \nodata	 & $0.11\pm0.02$ & $1.18\pm0.02$  & $104.83\pm6.71$ & $0.65\pm0.16$   & $2925\pm155$	\\
2008br	& IC 2522	&	3019	& $54579.2\pm1.6$	& $-15.30\pm0.2$ & \nodata	 & $0.45\pm0.02$ & \nodata	  & \nodata	    & $0.40\pm0.06$   & $4571\pm205$	\\
2008gr	& IC 1579	&	6831	& $54801.8\pm0.9$	& $-17.95\pm0.1$ & \nodata	 & $2.01\pm0.01$ & \nodata	  & \nodata	    & $0.17\pm0.03$   & $8731\pm521$	\\
2008if	& MCG -01-24-10 &	3440	& $54850.0\pm0.4$	& $-18.15\pm0.2$ & $4.03\pm0.07$ & $2.10\pm0.02$ & \nodata	  & $75.85\pm5.83$  & $0.08\pm0.03$   & $8685\pm441$	\\
2008in	& NGC 4303	&	1566	& $54851.5\pm0.8$	& $-15.48\pm0.5$ & $1.82\pm0.20$ & $0.83\pm0.02$ & \nodata	  & $92.20\pm6.71$  & $0.23\pm0.06$   & $6903\pm416$	\\
2008K	& ESO 504-G5	&	7997	& $54513.0\pm1.2$	& $-17.45\pm0.1$ & \nodata	 & $2.72\pm0.02$ & $2.07\pm0.26$  & $87.11\pm5.00$  & $0.16\pm0.07$   & $7954\pm511$	\\
2008M	& ESO 121-26	&	2267	& $54508.9\pm0.6$	& $-16.75\pm0.3$ & \nodata	 & $1.14\pm0.02$ & $1.18\pm0.26$  & $75.34\pm9.49$  & $0.22\pm0.04$   & $6674\pm290$	\\
2009bu	& NGC 7408	&	3494	& $54940.1\pm1.0$	& $-16.05\pm0.2$ & $0.98\pm0.16$ & $0.18\pm0.04$ & \nodata	  & \nodata	    & $0.50\pm0.13$   & $5943\pm516$	\\
2009N   & NGC 4487      &       1034    & $54877.1\pm0.5$	& $-15.35\pm0.4$ & \nodata	 & $0.34\pm0.01$ & \nodata	  & $89.50\pm5.83$  & $0.41\pm0.10$   & $5348\pm599$	\\
\enddata 
\tablecomments{Measurements made of our sample of SNe as mentioned in section 2. The first three columns present the SNe name and the host galaxy information:
Host galaxy name and their recession velocities. From column 4 to column 9 the photometric measurements: $t_{tran}$ (in \textit{B}-band), 
\textit{M$_{max}$}, \textit{$s_{1}$}, \textit{$s_{2}$}, \textit{$s_{3}$}, and $Optd$ (in \textit{V}-band) are presented. In column 10 we present 
the $a/e$, followed by the $H_{\alpha}$ velocity. }
\end{deluxetable}
\clearpage
\end{landscape}

\clearpage
\begin{landscape}
\begin{deluxetable}{ccccccccccc}
\tiny
\centering
\setlength{\tabcolsep}{1pt} 
\tablecolumns{8} 
\tablewidth{0pc} 
\tablecaption{We present the Pearson's r-parameter which indicates the strength of the correlation, together with
in brackets the number of events, and probability of finding such correlation by chance.}
\tablehead{ \colhead{\nodata} & \colhead{$a/e$} & \colhead{H$_{\alpha}$ vel.} &  \colhead{\textit{$s_{1}$}}
&  \colhead{\textit{$s_{2}$}} &  \colhead{\textit{$s_{3}$}} &  \colhead{M$_{max}$} & \colhead{Opt. thick}&}
\startdata
$a/e$ 		 & \nodata 			     & $-0.60$ (52; $2.05\times10^{-6})$ & $-0.74$ (23; $4.63\times10^{-5}$)  
		 & $-0.65$ (52; $1.63\times10^{-7})$ & $-0.64$ (15; 0.01)	         & $-0.53$ (42; $2.82\times10^{-4})$  
		 & $0.56$ (40; $1.90\times10^{-4})$ \\
H$_{\alpha}$ vel.& $-0.60$ (52; $2.05\times10^{-6})$ & \nodata 				 & $0.34$ (23; 0.11) 
		 & $0.64$ (52; $2.39\times10^{-7})$  & $0.51$ (15; 0.05)		 & $-0.77$ (42; $2.83\times10^{-9}$) 
		 & $-0.19$ (40; 0.23) \\
\textit{$s_{1}$} & $-0.74$ (23; $4.63\times10^{-5}$) & $0.34$ (23; 0.11) 		 & \nodata 
		 & $0.74$ (23; $5.02\times10^{-5})$  & $0.94$ (4; 0.06)	         	 & $0.54$ (19; 0.02)
		 & $-0.73$ (17; $8.10\times10^{-4})$ \\
\textit{$s_{2}$} & $-0.65$ (52; $1.63\times10^{-7})$ & $0.64$ (52; $2.39\times10^{-7})$  & $0.74$ (23; $5.02\times10^{-5})$ 
		 & \nodata 			     & $0.54$ (15; 0.40)		 & $0.70$ (42; $2.02\times10^{-7})$ 
		 & $-0.46$ (40; $2.80\times10^{-3})$ \\
\textit{$s_{3}$} & $-0.64$ (15; 0.01)                & $0.51$ (15; 0.05)		 & $0.94$ (4; 0.06)
		 & $0.54$ (15; 0.40)		     & \nodata 				 & $0.48$ (14; 0.08)
		 & $-0.72$ (15; $2.60\times10^{-3})$  \\
M$_{max}$        & $-0.53$ (42; $2.82\times10^{-4})$  & $0.77$ (42; $2.83\times10^{-9}$) & $0.54$ (19; 0.02)
		 & $0.70$ (42; $2.02\times10^{-7})$ & $0.48$ (14; 0.08) & \nodata 
		 & $-0.28$ (31; 0.12) \\
\textit{OPTd}    & $0.56$ (40; $1.90\times10^{-4})$  & $-0.19$ (40; 0.23)		 & $-0.73$ (17; $8.10\times10^{-4})$ 
		 & $-0.46$ (40; $2.80\times10^{-3})$ & $-0.72$ (15; $2.60\times10^{-3})$ & $-0.28$ (31; 0.12)
		 &  \nodata \\
\enddata
\end{deluxetable}
\clearpage
\end{landscape}

\end{document}